\documentclass[prd,preprint,tightenlines,floatfix,showpacs,preprintnumbers,nofootinbib,eqsecnum]{revtex4}
 \usepackage[dvips,final]{graphicx}
  \usepackage{amssymb,amsmath,epsfig,bm,pifont}
   \usepackage{pstricks,pst-node,pst-text,pst-3d}
\usepackage{color}

   \definecolor{mGreen}{rgb}{0.04,0.5,0.1}
   
    \newrgbcolor{DarkGreen}{0.04 0.5 0.1}
\def\be{\begin{equation}}
\def\ee{\end{equation}}
\def\ba{\begin{eqnarray}}
\def\ea{\end{eqnarray}}

\newcommand{\A}{{\mathcal{A}}}

\newcommand{\Ds}{\displaystyle}                            %%%%%%%%%

\def\as{\relax\ifmmode \alpha_s\else{$ \alpha_s${ }}\fi}  %%%%%%%%%
\def\abar{\relax\ifmmode{\bar{a}}\else{$\bar{a}${ }}\fi}  %%%%%%%%%
  \def\ie{\hbox{\it i.e.}{ }} %%%%%%%%%
   \def\eg{\hbox{\it e.g.}{ }}  %%%%%%%%%
\usepackage{graphics}
\usepackage{graphicx}% Include figure files
\usepackage{dcolumn}% Align table columns on decimal point
\usepackage{bm}% bold math
\usepackage{epsfig}
\usepackage{graphicx}
\usepackage{multirow}
\usepackage{dcolumn}% Align table columns on decimal point
\usepackage{graphicx,epsfig}%
\usepackage[symbol]{footmisc}
\usepackage{perpage}
\MakePerPage[2]{footnote}
\begin{document}
\title{
How to perform QCD analysis of DIS % eep inelastic scattering \\
in Analytic Perturbation Theory
}
\author{C\'esar Ayala$^{1,2}$\footnote{c.ayala86@gmail.com}}
 %\author{Alexander Bakulev$^2$ \footnote{deceased}}
 \author{S.~V.~Mikhailov$^2$ \footnote{mikhs@theor.jinr.ru}}

\affiliation{$^1$Department of Theoretical Physics and IFIC,
University of Valencia and CSIC, E-46100, Valencia, Spain\\
$^2$Bogoliubov Laboratory of Theoretical Physics, JINR, 141980 Dubna, Russia}

\date{\today}

\begin{abstract}
We apply (Fractional) Analytic Perturbation Theory (FAPT) to the QCD analysis
of the nonsinglet nucleon structure function $F_2(x,Q^2)$ in deep inelastic scattering up
to the next leading order
and compare the results with ones obtained within the standard perturbation QCD.
Based on a popular parameterization of the corresponding parton distribution
we perform the analysis within the Jacobi Polynomial formalism and under the control of
the numerical inverse Mellin transform.
To reveal the main features of the FAPT two-loop approach, we consider a wide range of momentum transfer
from high $Q^2\sim 100~{\rm GeV}^2$ to low $Q^2\sim 0.3~{\rm GeV}^2$ where the approach still works.
\end{abstract}

\maketitle

\section{I\lowercase{ntroduction}}
\label{sec:intr}
% \baselinestretch}{1.2}
QCD analysis of deep-inelastic scattering (DIS) data %in a wide range of variables  $(x, Q^2)$
provides one with new knowledge of hadron physics and serves as a test of reliability of our theoretical
understanding of the hard scattering of leptons and hadrons.
At large momentum transfer $q,~-q^2= Q^2 \gg 1$ GeV$^2$ we have the reliable description of DIS
that is based on the twist expansion and ``factorization'' theorems.
 At small (moderate) transfer $ Q^2 \lesssim 1$ (a few) GeV$^2$ this QCD description  faces
 two main problems:
 (i) the  high twist corrections to the leading twist contribution become important but remains poorly known;
 (ii) perturbative QCD (pQCD) becomes unreliable due to the fact that the QCD running coupling $\alpha_s(Q^2)$  grows and
 ``feels'' infra-red Landau singularity
    appearing  at the scale $Q \sim \Lambda_\text{QCD} \sim $ of a few tenth of GeV.
We discuss in this paper a solution of the last problem by applying to DIS analysis
a \textit{nonpower perturbative theory} whose couplings have no singularity at $Q^2 > 0$
and whose corresponding series possess a better convergence at low $Q^2$.

A widely used approach to resolve the aforementioned
problem is to apply the Analytic Perturbation Theory (APT) developed by Shirkov, Solovtsov {\it et al.\/}
\cite{ShS,MS,MSS,MSa,Sh1}.
There, the running QCD coupling
$a_s(Q^2) \equiv \alpha_s(Q^2)/4\pi$ of pQCD is transformed into an analytic (holomorphic)
 function of $Q^2$, $a^{1}_s(Q^2) \mapsto \A_1(Q^2)$, APT coupling.
This was achieved by keeping in the dispersion relation the spectral density
$\rho_1^{\rm (pt)}(\sigma) \equiv {\rm Im} \; a_s(Q^2=-\sigma - i \epsilon)/\pi$
unchanged on the entire negative axis in the complex $Q^2$-plane
(i.e., for $\sigma \geq 0$),  and setting it equal to zero along the unphysical cut
$0 < Q^2 < \Lambda^2$.
In the framework of APT the images $\mathcal{A}_n(Q^2)$ of integer powers
of the originals $a_s^n(Q^2)$, $a^n_s \mapsto \A_n$ following the same
dispersion relations were also constructed.
At low $Q^2$  the couplings $\A_n(Q^2)$  change slowly with $Q^2$ in contrast with
the original $a_s^n(Q^2)$ behaviour while at high $Q^2$ $\A_n(Q^2) \to a_s^n(Q^2)$.
Later, the correspondence  $a^\nu_s \mapsto \A_\nu$ was extended to noninteger powers/indices
$\nu$ in~\cite{KS06,BMS1,BMS2,BMS3,Bakulev} and was called Fractional APT (FAPT),
 which provides the basis for application to DIS.
In this respect let us mention a recent papers \cite{Sidorov2013}
where the processing of the DIS data  has been
performed in FAPT in the one-loop approximation
and the reasonable results for hadron characteristics has been obtained.

Various analytic QCD models can be constructed,
and have been proposed in the literature, among them in
Refs.~\cite{Nesterenko,Nesterenko2,Alekseev:2005he,Srivastava:2001ts,Webber:1998um,CV1,CV2}.
These models fulfill certain
additional constraints at low and/or at high $Q^2$.
For further literature on various analytic QCD models,
we refer to review articles
\cite{Prosperi:2006hx,Cvetic:2008bn,Bakulev}.
Some newer constructions of analytic models in QCD of $\A_1(Q^2)$
include those based on specific classes of $\beta$ functions with
nonperturbative contributions \cite{Belyakova:2010iw}
or without such contributions \cite{CKV1,CKV2,CCKKO}
and those based on modifications of the the spectral density
$\rho_1^{\rm (pt)} \mapsto \rho_1$
 $\left[\rho_1[\sigma]  \equiv {\rm Im} \; \A_{1}(Q^2=-\sigma - i \epsilon)/\pi\right]$
at low (positive) $\sigma$ where $\rho_1$ is parameterized in a specific manner
by adding two positive delta functions to $\rho_1^{\rm (pt)}$, cf.~\cite{CCEMAyala}.

The possibility to extend the DIS analysis formally in the whole $Q^2$ range together with
the effect of slowing-down of the FAPT evolution of the parton distribution functions (PDF) in the
low $Q^2$ region are attractive phenomenological features of FAPT.
A number of works deal with this task in a naive form \cite{naiveF2},
where the authors show that at very low $Q^2$ and Bjorken variable $x$ APT agrees with experimental data.
Besides, the applicability of the APT approach was analyzed in the Bjorken polarized sum rule \cite{BjSR}
confirming that the range of validity of APT is down to $Q\sim\Lambda_{\rm QCD} \simeq350$ MeV,
as compared to experimental data.
The common feature of these works was taking into consideration some nonperturbative effects
against the background of APT, i.e.,
higher twists in \cite{naiveF2,BjSR,OT2013} or an effective constant gluon mass in \cite{GluonProp}.

The basis for applying FAPT to low energies in this approach
is  the factorization theorem that allows one to shift the frontier
between the perturbative and nonperturbative effects via the variation of the factorization scale.
Therefore, we shift the range where perturbation series is applicable in FAPT, as it was demonstrated in
\cite{BjSR} (see reviews of this issue in \cite{Bakulev},
where this phenomenon was also discussed for pion form factors).

Our goal here is to elaborate a general scheme of DIS data processing in the framework of FAPT
taking as a pattern the DIS analysis at NLO.
In this respect the discussion here can be considered as an extension of the
partonic results of the article \cite{Sidorov2013} on the higher-loop level.
We shall focus on the specifics of the Dokshitzer-Gribov-Lipatov-Altarelli-Parisi (DGLAP)
evolution of the PDF $f(x;\mu^2)$  in FAPT.
We involve into consideration the coefficient function $C(x,a_s)$ of the process
and compare the final result with a similar one in pQCD.
An important problem of higher twist contribution remains untouched here,
but higher twist effects can be taken as an unknown function $h(x)$,
i.e. $h(x)/Q^2$ \cite{Sidorov2013}, or as a constant \cite{BjSR} $\mu_4/Q^2$,
or as an effective sum of all twists contributions in \cite{OT2013}.
We stress that HT effects are only indirectly affected by the analytization procedure.
The behavior
of HT will be given by the fit of experimental data together with the
corresponding parton distribution functions \cite{Sidorov2013,naiveF2,BjSR}.
Besides, in \cite{BjSR} the authors included more terms in the HT expansion
and demonstrated that they are essentially smaller and quickly decreasing.
Because of this (theoretically) unknown behavior
we avoid this problem since we pretend to provide perturbation tools  how to deal
with FAPT, while the pure phenomenological analysis
is transferred to future investigations.

Let us recall that the DIS analysis can be performed  in a few  different ways:
one of them is provided by the Mellin moments defined via inelastic structure functions (SFs) $F(x,Q^2)$,
 \begin{equation}
M(n,Q^2)=\int_0^{1}dx x^{n-1}F(x,Q^2),
\qquad (n=1,2,3, ...) \ .
\label{moments}
\end{equation}
The second approach is based on the direct application of the DGLAP
integro-differential evolution equations~\cite{dglap}
to PDF $f$,
 while the observable SF is the Mellin convolution of the coefficient function and PDF, $F= C\ast f$.
The third approach makes use of the Jacobi Polynomial expansion method~\cite{Krivokhizhin}.
Just this method will be used in this work.

The paper is organized as follows:
in Sec.~\ref{TB}, we present a theoretical background where we describe the
Jacobi Polynomial (JP) method and how to calculate  free parameters in order to obtain
the nonsinglet structure function.
In Sec.~\ref{FAPT}, we briefly describe the FAPT approach and
derive the DGLAP evolution for the moments $M(n,Q^2)$ in FAPT.
We present in Sec.~\ref{NR} the free parameters obtained in the
analysis of the so called MSTW parameterization, see \cite{Martin:2009iq} for details,
and the nonsinglet SFs at LO.
Section~\ref{NR} contains the results of the analysis of numerical realization of the
FAPT evolution and
the comparison with the results of analogous calculations in pQCD.
Finally, in Sec.~\ref{sec:summ} we summarize our conclusions.
Important technical details including new findings are collected in four appendices.

\section{J\lowercase{acobi} P\lowercase{olynomial} \lowercase{expansion for} DIS \lowercase{analysis}}
\label{TB}
We shall focus  here on nonsinglet (NS) structure functions, $F_\text{NS}(x,Q^2)$,
with their corresponding Mellin moments  $M_\text{NS}(n,Q^2)$ (via Eq.~(\ref{moments}))
to avoid technical complications of the coupled system solution in the singlet case.
The PDFs $f_p(x,\mu^2)$ are universal process-independent
densities explaining how the whole hadron momentum $P$ is partitioned in
$x\cdot P$, i.e., the momentum carried by the struck
parton (see, for instance~\cite{Buras:1979yt}).
The $x$-dependence of PDF
is formed at a hadron scale of an order of $P^2$ by nonperturbative forces,
while its dependence on factorization/renormalization scale $\mu^2$
can be obtained within perturbation theory.

A brief description of the evolution of the Mellin moments in pQCD, up to NLO, is outlined in Appendix~\ref{app1} as well as the theoretical background with our notation and conventions.
We consider  the scale $Q_0^2$ as a reference scale for the solution of the evolution equation~(\ref{mns})
where the PDFs are regarded as functions of $x$ and the parameters are fixed by comparison with DIS data.
In particular, we use here the data-based MSTW PDFs (see~\cite{Martin:2009iq}, where
$Q_0^2=1~{\rm GeV}^2$).
Namely,
\ba
x u_v(x,Q_0^2)&=&A_u x^{\eta_1}(1-x)^{\eta_2}(1+\epsilon_u \sqrt{x}+\gamma_u x),\\
x d_v(x,Q_0^2)&=&A_d x^{\eta_3}(1-x)^{\eta_4}(1+\epsilon_d \sqrt{x}+\gamma_d x),
\label{mstw}
\ea
where the values of $A_{u,d}, \eta_k$ ($k=1,...,4$), $\epsilon_{u,d}$ and $\gamma_{u,d}$ can be
found in~\cite{Martin:2009iq}. We use only the valence quark PDFs because
the NS PDF $f_{\rm NS}$ can be expressed as $f_{\rm NS}(x,Q^2)=u_v(x,Q^2)-d_v(x,Q^2)$
(see Appendix~\ref{app1} for details).
 The NS SF $F_2(x)=\left(C\ast f_{\rm NS}\right)(x) $
is represented as the Mellin convolution of coefficient function $C$ of the process
and the corresponding PDF $ f_{\rm NS}$.
The $F_2(x)$ can be expanded in the Jacobi Polynomials
$\Theta_n^{\alpha \beta}(x)$, which was developed in Refs.~\cite{Krivokhizhin},
in truncating the expansion at $n=N_{max}$,
where the method converges (see, for review \cite{Kotikov}):
 \begin{equation}
F_2(x,Q^2;N_{max})=\omega^{\alpha \beta}(x)\sum_{n=0}^{N_{max}}\Theta_n^{\alpha \beta}(x)
\sum_{j=0}^n C_j^{(n)}(\alpha,\beta)M_\text{NS}(j+1,Q^2).
\label{jpSF}
\end{equation}
Here $M_\text{NS}(n,Q^2)$ are the Mellin moments of nonsinglet SF calculated explicitly in Eq.~(\ref{mns});
~$\omega^{\alpha \beta}(x)=x^\alpha(1-x)^\beta$ is the weight function and the parameters
$\alpha,\beta$ will be obtained by fitting to the data.
The Jacobi Polynomials $\Theta_n^{\alpha\beta}(x)$
are defined as an expansion series by means of
 \begin{equation}
\Theta_k^{\alpha \beta}(x)=\sum_{j=0}^k C_j^{(k)}(\alpha,\beta)x^j.
\label{jpdef}
\end{equation}
They satisfy the orthogonality relation
 \begin{equation}
\int_0^1\omega^{\alpha\beta}(x)\Theta_k^{\alpha \beta}(x)\Theta_l^{\alpha \beta}(x)=\delta_{kl}.
\label{orthogonalJP}
\end{equation}

Another way to obtain  SF $F_2(x,Q^2)$ is to take
the inverse Mellin transform $\mathcal{M}^{-1}$ under the moments $M_\text{NS}(n,Q^2)$
(i.e., the inverse of Eq.~(\ref{moments})).
Choosing a convenient path of integration one obtains for $F_2$
\begin{equation}
F_2(x,Q^2)\equiv \mathcal{M}^{-1}\left\{M_\text{NS}(n,Q^2) \right\}=\frac{1}{2\pi i}\int_{c-i\infty}^{c+i\infty}
x^{-n}M_\text{NS}(n,Q^2)dn,
\label{exInvM}
\end{equation}
here we take the path  along a vertical line  ${\rm Re}(n)=c$.
 We perform the ``exact''
numerical Inverse Mellin Transform, further comparing the results
with the Jacobi Polynomial method,
only at the one-loop level due to technical limitations.
In this way, we estimate the accuracy of the applied polynomial  method,
the results of this numerical verification are outlined in Appendix \ref{app3}.

Let us finally mention that one can take SF $F_3$ instead of the NS $F_{2}$  to consider,
\eg, the neutrino DIS results of the CCFR collaboration, like it was started in \cite{Sidorov2013}.
 This replacement will lead to only minor changes of technical details in
 the procedure elaborated below.

 \section{Fractional Analytic Perturbation Theory \lowercase{and} DIS}
\label{FAPT}
It is known that the perturbative QCD coupling suffers from unphysical (Landau) singularities
at $Q^2\sim\Lambda^2$.
This prevents the application of perturbative QCD in the low-momentum spacelike regime
and, in part, impedes the investigation of high twists in DIS.
Our goal here is not to discuss the motivation and complete construction of FAPT,
which couplings $\A_\nu$ are free of the aforementioned problems,
but present to reader illustrations of the properties of this nonpower perturbation theory
that are important for further DIS analysis.
\subsection{E\lowercase{lements of} FAPT}
Application of the Cauchy theorem to the running coupling
$a_s^\nu(Q^2)\equiv\left(\alpha_s(Q^2)/4\pi\right)^\nu $,
established in \cite{ShS,MS,MSS,MSa,Sh1} and developed in \cite{KS06,BMS1,BMS2,BMS3,Bakulev},
 gives
us the following dispersion relation (or K\"{a}ll\'en-Lehmann spectral representation)
for the images $\A_\nu^{(l)}$ in the spacelike domain
\begin{equation}
 \A_\nu^{(l)}(L) =\int_0^{\infty} \frac{\rho_\nu^{(l)}[\sigma]}{\sigma+Q^2}d\sigma
 =
 \int_{-\infty}^{\infty} \frac{\rho_\nu^{(l)}(L_\sigma)}{1+{\rm exp}(L-L_\sigma)}dL_\sigma\,
, \label{AnuSD}
 \end{equation}
(where~$L_\sigma=\ln(\sigma/\Lambda^2)$) that has no unphysical (Landau) singularities.
For the timelike regime  analogous coupling reads
 \begin{equation}
 \mathfrak{A}_\nu^{(l)}(L_s)=\int_s^{\infty} \frac{\rho_\nu^{(l)}[\sigma]}{\sigma}d\sigma
 =\int_{L_s}^{\infty} \rho_\nu^{(l)}(L_\sigma) dL_\sigma.
 \label{UnuSD}
 \end{equation}
Here, $\A_\nu^{(l)}(L) $ is the FAPT image of the
QCD coupling $a_{s(l)}^\nu(L)$  in the Euclidean (spacelike) domain with $L=\rm{ln}(Q^2/\Lambda^2)$ and the label
$l$ denotes running in the $l$-loop approximation,
whereas in the Minkowski (timelike) domain,
we used  in (\ref{UnuSD}) $L_s={\rm ln}(s/\Lambda^2)$.
It is convenient to use the following representation for the spectral densities $\rho_\nu^{(l)}$:
\ba
&&\rho_\nu^{(l)}(L_\sigma) \equiv \frac{1}{\pi}{\rm Im} \left(a_{s(l)}(L-i\pi)\right)^\nu
=\frac{{\rm sin}[\nu\varphi_{(l)}(L)]}{\pi\left( R_{(l)}(L) \right)^\nu}, \label{eq:rho}\\
&&R_{(l)}(L)= \big|a_{s(l)}(L-i\pi) \big|;~~\varphi_{(l)}(L)=arg\left(a_{s(l)}(L-i\pi)\right). \nonumber
\ea
From the definition (\ref{AnuSD}) and Eq.(\ref{eq:rho}) it follow that there is no
standard algebra for the images $\A_\nu$, \ie $\A_\nu\A_\mu \neq \A_{\nu+\mu} $ that justifies the name
\textit{nonpower perturbative theory}.

In the one-loop approximation, the $\varphi_{(1)},~R_{(1)}$ has the simplest form, i.e.,
\begin{equation}
\varphi_{(1)}(L)={\rm arccos} \left(\frac{L}{\sqrt{L^2+\pi^2}} \right),
\quad
R_{(1)}(L)=\beta_0 \sqrt{L^2+\pi^2}.\label{eq:rho-1}
\end{equation}
Substituting Eq.(\ref{eq:rho-1}) in Eq.(\ref{eq:rho}) for $\rho_1^{(1)}$ and then the result $\rho_1^{(1)}(L_\sigma)$
 in Eq.(\ref{AnuSD}), one reproduces at $Q^2=0$ the well-known expression for maximum value of $\A_1^{(1)}(L)$, $ \A_1^{(1)}(L=-\infty) $ \cite{ShS},
\begin{equation}
 \A_1^{(1)}(-\infty)=\int_{-\infty}^{\infty} \frac{dL_\sigma}{\beta_0(L_\sigma^2+\pi^2)} = \frac1{\beta_0} >  \A_1^{(1)}(L)\,.
 \end{equation}
\textbf{ }
At the two-loop level, they have a more complicated form.
To be precise, one gets
\be
a_{s(2)}=-\frac{1}{c_1}\frac{1}{1+W_{-1}(z_W(L))},
\ee
and
\ba
R_{(2)}(L) &=& c_1(n_f)\left| 1+W_{-1}(z_W (L+i\pi))\right|,
\nonumber\\
\varphi_{(2)}(L)&=& {\rm arccos}\left[\frac{{\rm Re}\left(1+W_{-1}(z_W (L+i\pi))\right)}{R_{(2)}(L)} \right],
\ea
with $W_{-1}(z)$ being the appropriate branch of the Lambert function, $z_W(L)=-c_1^{-1}(n_f)
e^{-1-L/c_1(n_f)}$,
 $c_k(n_f)\equiv \beta_k(n_f)/\beta_0(n_f)^{k+1}$, where $\beta_k$ are the QCD $\beta$-function
 coefficients
  and $n_f$ is the number of active quarks, see the expressions in Appendix~\ref{app0}.
For our purpose we use here only the two-loop couplings like $a^{\nu}_{(2)s},~\A_\nu^{(2)}$.
Extensions up to four-loops
can be found in ~\cite{Bakulev:2012sm}.

Now we  implement this formalism with the help of numerical calculation
 with the main module Mathematica package FAPT.m of ~\cite{Bakulev:2012sm}
(confirmed by a recent program in ~\cite{Ayala2014}).
According to this and using the corresponding notation from \cite{Bakulev:2012sm} in the RHS of
Eqs.(\ref{AnuFAPTprog}-\ref{alphaFAPTprog}),  we have
\ba
 \A_\nu^{(l)}(L)&=& \frac{{\rm AcalBar}l [L,n_f,\nu]}{(4\pi)^\nu},
 \quad
 (l=1\div 4; n_f=3 \div 6)
 \label{AnuFAPTprog}
\\
 \mathfrak{A}_\nu^{(l)}(L_s)&=& \frac{{\rm UcalBar}l [L,n_f,\nu]}{(4\pi)^\nu},
 \quad
 (l=1\div 4; n_f=3 \div 6)
  \label{UnuFAPTprog}
 \ea
For the coupling in  pQCD we obtain
\begin{equation}
a_{s(l)}(L={\rm ln}(Q^2/\Lambda^2))=\frac{\alpha{\rm Bar}l[Q^2,n_f,\Lambda]}{4\pi},
\quad
 (l=1\div 4).\label{alphaFAPTprog}
\end{equation}

The correspondence between the pQCD expansion and FAPT one is based on the
linearity of the transforms in Eqs.(\ref{AnuSD}) and (\ref{UnuSD}), see \cite{Sh1}.
This can be illustrated for the simple case of a single scale quantity $D(Q^2,\mu^2_R)$,
calculated within minimal subtraction renormalization schemes and taken at the renormalization scale $\mu^2_R=Q^2$.
The expansions for $D$ and for its image $D \mapsto {\cal D}$ are written as
\ba
\text{pQCD:}~D(Q^2)     &=&d_0a_s^{\nu}(Q^2)~+ \sum_n d_n~a_s^{n+\nu}(Q^2)   \nonumber \\
\text{FAPT:}~{\cal D}(Q^2)&=&d_0\A_{\nu}(Q^2)+ \sum_n d_n~\A_{(n+\nu)}(Q^2) \label{eq:DtoD}
\ea
at the \textit{same coefficients $d_i$} that are numbers at $\mu^2_R=Q^2$.

\subsection{ FAPT for DGLAP evolution in NLO approximation}
We start with the well-known solution of DGLAP equation for the nonsinglet PDF $f_\text{NS}$ in NLO approximation.
This solution is combined with the corresponding coefficient function $C(x,a_s)$ -- the parton cross-section
taken at the parton momentum $xP$.
This is presented in Appendix \ref{app1}
in the form of Eq.(\ref{mns}) for the moments $M_\text{NS}$ of the NS SF $F_2$.

Rewriting Eq.~(\ref{mns}) in the approximate form, i.e., neglecting the $\mathcal{O}(a_s^2)$ terms
in the two-loop evolution factor, one arrives at the commonly used relation
\begin{equation}
M_\text{NS}(n,Q^2)=\frac{a_{s(2)}^{d_\text{NS}(n)}(Q^2)+
\left(C_\text{NS}^{(1)}(n)+\frac{\beta_1}{\beta_0}p(n) \right)a_{s(2)}^{d_\text{NS}(n)+1}(Q^2)}
{a_{s(2)}^{d_\text{NS}(n)}(Q_0^2)+
\left(C_\text{NS}^{(1)}(n)+\frac{\beta_1}{\beta_0}p(n) \right)a_{s(2)}^{d_\text{NS}(n)+1}(Q_0^2)}
 ~M_\text{NS}(n,Q_0^2).
 \label{appMpQCD}
\end{equation}
The use of  FAPT will change in this scheme the sense of expansion parameters $a_{s}$ in accordance with (\ref{eq:DtoD}).
An analogous  evolution relation for the analytic images of the moments $M_\text{NS}$,
$M_\text{NS}\mapsto \mathcal{M}_\text{NS}$, can be obtained from Eq.(\ref{appMpQCD}) by
replacing the powers $(a_s)^\nu$ with the FAPT couplings $\A_\nu$
(with $\nu$ being here an index rather than a power) \cite{BMS2}
 and reads
\begin{equation}
{\cal M}_\text{NS}(n,Q^2)=\frac{ \A_{d_\text{NS}(n)}^{(2)}(Q^2)+
\left(C_\text{NS}^{(1)}(n)+\frac{\beta_1}{\beta_0}p(n) \right)\A_{d_\text{NS}(n)+1}^{(2)}(Q^2)}
{ \A_{d_\text{NS}(n)}^{(2)}(Q_0^2) +
\left(C_\text{NS}^{(1)}(n)+\frac{\beta_1}{\beta_0}p(n) \right) \A_{d_\text{NS}(n)+1}^{(2)}(Q_0^2) }
~{\cal M}_\text{NS}(n,Q_0^2).
 \label{mnsapt}
\end{equation}
The implementation of the proposed calculation in the form of (\ref{AnuFAPTprog},\ref{alphaFAPTprog})
is quite direct.
The FAPT evolution relation (\ref{mnsapt}) for the moments is the main result of the Section.
Further, we shall use code (\ref{AnuFAPTprog}) from \cite{Bakulev:2012sm} to obtain
$\A_\nu^{(2)}(L), \left(a_{s(2)}^{\nu}(L)\right)$
numerically.

The last approximation was taken up to $\mathcal{O}(\A_{d_\text{NS}+1})$
since the contribution of the next term in the FAPT expansion in Eq.(\ref{mnsapt})
is negligible in comparison  with the previous one
(as we demonstrate in Appendix~\ref{app2}).
This analytic version of the moment evolution does not face  any problems
at low energies due to the boundedness of couplings and rapid convergence of
the FAPT series.

In the absence of a fit of experimental data for the FAPT model we propose a relation
for the initial moments at $Q^2_0$:
\ba
f_\text{NS}(n,Q^2_0)&=&\frac{M_\text{NS}(n,Q_0^2)}{a_s(Q_0^2)^{d_\text{NS}(n)}+
\left(C_\text{NS}^{(1)}(n)+\frac{\beta_1}{\beta_0}p(n) \right)a_s(Q_0^2)^{d_\text{NS}(n)+1}}
\nonumber\\
&=&\frac{{\cal M}_\text{NS}(n,Q_0^2)}{\A_{d_\text{NS}(n)}^{(2)}(Q_0^2) +
\left(C_\text{NS}^{(1)}(n)+\frac{\beta_1}{\beta_0}p(n) \right) \A_{d_\text{NS}(n)+1}^{(2)}(Q_0^2)},
\label{inMom}
\ea
where the moment of PDF (see Eq.(\ref{eq:NSmpdf})) in pQCD stands in the LHS,
 while the moment for PDF in FAPT stands in the RHS of the second equation.
In other words, we take the same initial PDF as in pQCD from the MSTW data for these both cases
(in \cite{Sidorov2013} the parameters were taken the same since the difference
between them was negligible).
We can use either the Jacobi Polynomial expansion or directly the inverse Mellin transform
( Appendix \ref{app3}).

\section{ R\lowercase{esults of numerical analysis}}
\label{NR}
The accuracy of the SF approximation by a finite number of Jacobi Polynomials (truncated at $N_{max}$) depends on
the choice of the weight-function parameters.
Therefore, we test the nonsinglet SF,
given by the MSTW data, by searching for the minimum of ($Q^2=Q_0^2$):
 \begin{equation}
\chi_{\alpha,\beta}^2=\left| F_2^{(theor),N_{max}}/F_2^{(exp)}-1\right|^2,
\label{chi2}
\end{equation}
where we have used Eqs.~(\ref{moments}) and~(\ref{mns}) at $Q^2=Q_0^2$.
Thus, we have
$F_2(x,Q_0^2)\equiv F_2^{(exp)}(x,Q_0^2)$ and from Eq.~(\ref{jpSF})
$F_2^{(theor),N_{max}}(x,Q_0^2)\equiv F_2^{N_{max}}(x,Q_0^2)$.
 Then, we determine the values of $\alpha$ and $\beta$ that provide the best fit to the data
 for different values of $N_{max}$.
 At the one loop level we find: $N_{max}=13$, $\alpha=0.05$, and $\beta=3.03$ for
 $\chi^2\approx 10^{-9}$, whereas for two loops we get (for even PDFs only):
 $N_{max}=13$, $\alpha=-0.8$, and $\beta=2.99$ for $\chi^2\approx 10^{-9}$.
 To evolve nonsinglet moments, we need to fix the values of the QCD scale
 $\Lambda_{1,2}(n_f=3)$ in the leading and next-to-leading order,
 taken in~\cite{Martin:2009iq} from the comparison with data,
 where $\alpha_s^{(1loop)}(Q_0^2=1{\rm GeV}^2)=0.682 \Rightarrow \Lambda_1(n_f=3)=0.359~{\rm GeV}$
 and $\alpha_s^{(2loop)}(Q_0^2=1~{\rm GeV}^2)=0.491 \Rightarrow \Lambda_2(n_f=3)=0.402~{\rm GeV}$.
  In the case of FAPT, the scales $\Lambda_{1,2}^{\rm FAPT}(n_f=3)$ must be taken into account
  very carefully.
  The authors of \cite{Sidorov2013}  fixed the $\Lambda$ value directly from the comparison with
  the data in the leading order (where $Q_0^2=3~{\rm GeV}^2$) and obtained
  $\Lambda_{1}^{\rm FAPT}(n_f=4)=0.275\pm 0.039~{\rm GeV}$
  that corresponds to $\Lambda_{1}^{\rm FAPT}(n_f=3)=0.333\pm 0.050~{\rm GeV}$.
  We can see that the perturbative and the analytic values of $\Lambda$  are
  close to each other at least inside the margin of errors.
  For this reason, we will take
  $\Lambda_{1,2}(n_f=3) \simeq \Lambda_{1,2}^{\rm FAPT}(n_f=3)$ for simplicity
  (recalling that an appropriate value should be taken from the analysis
  of the experimental data but this goes beyond the scope of this work).
  The couplings in pQCD and in FAPT were calculated with the Mathematica package developed
  by Bakulev and Khandramai in~\cite{Bakulev:2012sm}
  where the heavy flavour thresholds were taken into account.

\begin{figure}[ht] %\unitlength=1mm
\centering\includegraphics[width=140.mm]{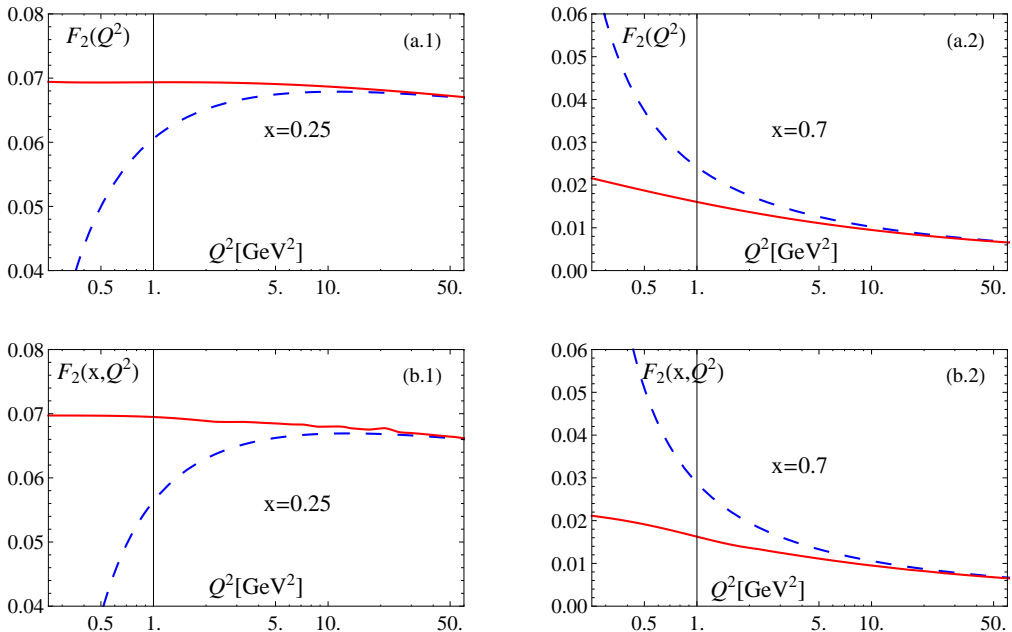}
%\centering\includegraphics[width=150mm]{graph1.pdf}
\vspace{-0.5cm}
\caption{\footnotesize Nonsinglet SF  $F_2(x,Q^2)$ vs $Q^2$ at (a) LO and (b) NLO. The Bjorken
$x=0.25$ for (a.1) and (b.1), and $x=0.7$ for (a.2) or (b.2).
 The solid line represents the FAPT results and the dashed line -- the pQCD ones.}
\label{figsf1}
 \end{figure}

\begin{figure}[ht] %\unitlength=1mm
\centering\includegraphics[width=160.mm]{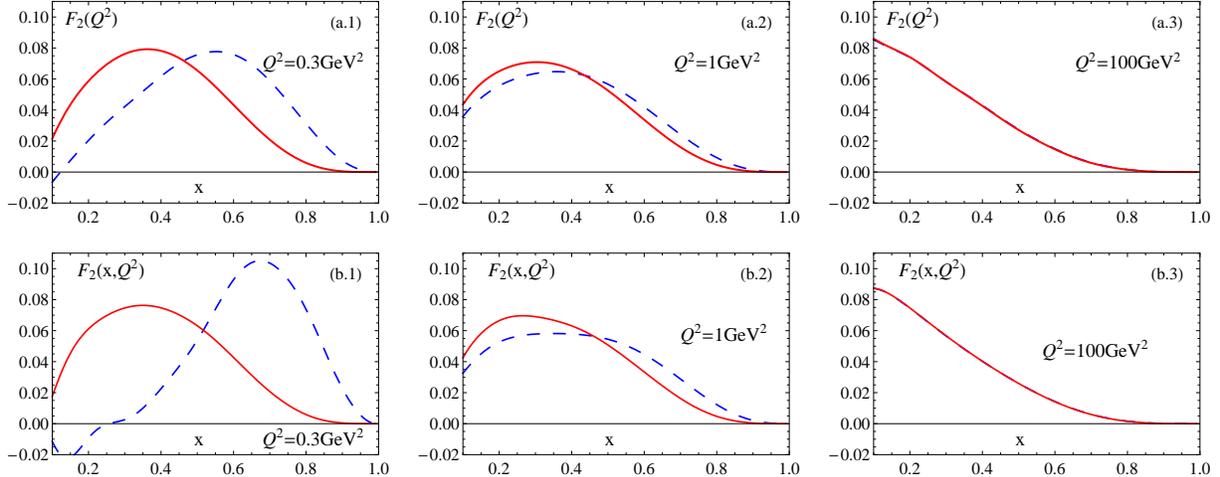}
%\centering\includegraphics[width=150mm]{graph1.pdf}
\vspace{-0.5cm}
\caption{\footnotesize Nonsinglet SF $F_2(x,Q^2)$ vs $x$ at (a) LO and (b) NLO.
The energy scale is $Q^2=0.3~{\rm GeV}^2$ in (a.1), (b.1), $Q^2=1~{\rm GeV}^2$ in (a.2), (b.2)
and $Q^2=100~{\rm GeV}^2$ in (a.3), (b.3). The solid line represents the FAPT results and the dashed line -- the
pQCD ones.}
\label{figsf2}
 \end{figure}

  Taking into account the above estimates of the initial parameters,
  we substitute Eqs.~(\ref{appMpQCD}) and~(\ref{mnsapt})
  into Eq.~(\ref{jpSF}), and
  obtain the evolution of SFs up to NLO in pQCD or FAPT, respectively.
  We show the final results of the evolution in Figs.~\ref{figsf1},~\ref{figsf2}
  using for DIS the character interval $0.3\leq Q^2\leq 100~{\rm GeV}^2$.
 %\vspace{-0.3cm}
 In Fig.~\ref{figsf1} we fix $x$ at two different values:
  $x=0.25$ in (a.1), (b.1), and  $x=0.7$ in (a.2), (b.2),
  where (a) and (b) represent the LO and NLO results, respectively.
   In Fig.~\ref{figsf2}, we fix $Q^2$ at three different values: $Q^2=0.3~{\rm GeV}^2$ in (a.1), (b.1),
  with the initial point $Q^2=1~{\rm GeV}^2$ in (a.2), (b.2), and $Q^2=100~{\rm GeV}^2$ in (a.3), (b.3),
  where again (a) and (b) represent the LO and NLO, respectively.

% \begin{table}
% \caption{\label{tab1} The best fit for weight function parameters $\alpha$ and $\beta$
% minimizing $\chi^2$ for diferents values of $N_{max}$.}
% \begin{tabular*}{0.75\textwidth}{@{\extracolsep{\fill}} | l | c | c | r |}\hline
%  $N_{max}$ & $\chi^2$ & $\alpha$ & $\beta$\\
% \hline \hline
% 3 & $10^{-3}$ & -0.2162 & 3.1595 \\
% 5 & $3.8\times10^{-6}$ & -0.8143 & 3.0072\\
% 7 & $6.6\times10^{-7}$ & -0.4963 & 3.0333 \\
% 9 & $1.8\times10^{-7}$ & -0.2674 & 3.0331 \\
% 13 & $1.6\times10^{-9}$ & 0.01136 & 3.0330 \\
% \hline
% \end{tabular*}
% \end{table}
%\newpage

\section{S\lowercase{ummary}}
\label{sec:summ}
The main goal of this work is to propose a new theoretical tool for the DIS analysis,
based on Fractional Analytic Perturbation Theory, to the DIS community.
This approach allows one to analyze formally the leading-twist structure function in the whole
$Q^2$ range. This conclusion is explicitly  shown in  Figs.~\ref{figsf1} and~\ref{figsf2}.
The scheme of the approach is formulated in Sec.~\ref{FAPT} and applied for data processing in Sec.~\ref{NR}.
Our consideration is restricted to the leading twist.
The higher twist contributions (HT) can be taken into account by a fit of experimental data
together with PDFs.
Moreover, the role of the stability of APT for this fit was pointed out in
\cite{Sidorov2013,naiveF2,BjSR, OT2013} (and in Introduction here).
Our investigation reveals the following main features of applying FAPT:
\begin{itemize}
\item
Structure function $F_2(x,Q^2)$ at fixed $x$ changes very slowly in the entire range of $Q^2$.
\item
At high $Q^2$ evolution ($Q^2\gtrsim 100~{\rm GeV}^2$) the pQCD and FAPT distributions become practically
equal.
\item
The evolution in FAPT is more gradual (i.e., it evolves slower) and smoother than in pQCD.
\item
The new analytic (FAPT) series converge faster than the pQCD series. From inspection of
Figs.~\ref{figsf1},~\ref{figsf2} it is obvious that the one- and the two-loop FAPT approximations
 do not differ significantly from each other
(the difference is less than $1\%$).\\
\end{itemize}

In this work, we have analyzed  only the nonsinglet part,
the consideration of the singlet part can be performed along the same line but requires more complicated
formulas and cumbersome numerical calculations.
This is the task for forthcoming  investigation.
Other important issues to complete this FAPT approach as the reliable tool for DIS
is to add the target mass corrections (TMC) and
the aforementioned HT contributions in our scheme of calculation.
These improvements will help one to clarify in future the behavior at very low energies ($Q^2\sim
0.3~{\rm GeV}^2$) in more detail.
It would be important to emphasize, that the FAPT approach admits investigation of the HT contributions
in the most sensitive regime of moderate/small $Q^2$ due to the high stability of the
radiative corrections.
\begin{acknowledgments}
\noindent
This investigation was started by the late A. P. Bakulev
to whom we dedicate this work.
We are grateful to G. Cveti\v{c} for useful comments and to N. G. Stefanis for
careful reading of the paper and  many valuable critical remarks.
We thank A.~V. Sidorov and O.~P. Solovtsova for the useful remarks.
This work was supported by the scientific program of the
  the Russian Foundation for Basic Research Grant No.\ 14-01-00647,
  BelRFFR--JINR grant F14D-007 (S.V.M), and by CONICYT Fellowship ``Becas
  Chile'' Grant No.74150052 (C.A).
\end{acknowledgments}

\appendix
\section{E\lowercase{xplicit expressions for} NLO \lowercase{$\beta$-functions,
anomalous dimensions and coefficient
functions of} DIS \lowercase{(nonsinglet case)}}
\label{app0}
The renormalization group equation for $\Ds a_s =\frac{\alpha_{s}(L)}{4 \pi}$ at
the expansion of the $\beta$-function up to the NLO
approximation is given by
\be
\label{eq:betaf}
 \frac{d}{d L} a_s(L)
  =-\beta(a_s(L))=
  - \beta_0 a_s^2(L)
    - \beta_1 a_s^3(L)+\ldots\,, \\
\ee
 where the first two beta coefficients are

\be
\beta_0 = \frac{11}{3}\,C_{\rm A} - \frac{4}{3}\,T_{\rm R} n_f
\,,~ \beta_1
=
  \frac{34}{3}\,C_{\rm A}^{2}
  - \left(4C_{\rm F}
  + \frac{20}{3}\,C_{\rm A}\right)T_{\rm R} n_f\,.
\label{eq:beta0&1}
\ee
The anomalous dimensions of composite operators in LO, $\gamma_\text{NS}^{(0)}(n)$, NLO
$\gamma_\text{NS}^{(1)}(n)$
and the coefficient function $C_\text{NS}^{(1)}(n)$ are expressed
by means of transcendental sums $S_{\alpha}(n)$, see, e.g., \cite{YndBook},
\begin{subequations}
\ba
 \label{eq:AD-LO-NLO}
\gamma_\text{NS}^{(0)}(n)&=&2C_F\left[1-\frac{2}{n(n+1)}+4\left(S_{1}(n)-1\right)\right],
\ea
\ba
\gamma_\text{NS}^{(1)\pm}(n)&=& \left(C_F^2-\frac{1}{2}C_F C_{\rm A} \right) \times
\left\{16S_1(n)\frac{2n+1}{n^2(n+1)^2}+16 \left[2S_1(n)-\frac{1}{n(n+1)} \right] \right.
\nonumber
\\
&&\left.
\cdot \left[ S_2(n)-S_2^{\pm}  \left(\frac{n}{2} \right)\right]
+ 64 \tilde{S}^{\pm}(n)+24S_2(n)-3-8S_3^{\pm}\left(\frac{n}{2} \right)
\right. \nonumber
\\
&&\left.-8\frac{3n^3+n^2-1}{n^3(n+1)^3}\mp 16\frac{2n^2+2n+1}{n^3(n+1)^3} \right\}
\nonumber
\\
&&
 +C_FC_{\rm A} \left\{ S_1(n)\left[\frac{536}{9}+8\frac{2n+1}{n^2(n+1)^2} \right] -16S_1(n)S_2(n)
\right.
 \nonumber
\\
&&
\left. +S_2(n)\left[-\frac{52}{3}+\frac{8}{n(n+1)} \right]
 -\frac{43}{6}-4\frac{151n^4+263n^3+97n^2+3n+9}
 {9n^3(n+1)^3} \right\}
 \nonumber
\\
&&
 +C_F N_F T_R \left\{-\frac{160}{9}S_1(n)+\frac{32}{3}S_2(n)+\frac{4}{3}+16\frac{11n^2+5n-3}{9n^2(n+1)^2}
 \right\}\,,
\label{AD}
\ea
\ba
C_\text{NS}^{(1)}(n)&=&C_F
\left(2S_1^2(n)+3S_1(n)-2S_2(n)-\frac{2S_1(n)}{n(n+1)}+\frac{3}{n}+\frac{4}{n+1}+\frac{2}{n^2}
-9\right)\,.
\ea
 \end{subequations}
On the other hand, the series $S_\alpha (n)=\sum_{k=1}^{n}\frac{1}{k^\alpha}$  can be expressed
via the generalized Riemann
$\zeta$ functions, see \cite{Erde53}, that are analytic functions in \textit{both variables} $\alpha, n$:
\begin{subequations}
 \label{eq:S}
\ba
S_1 (n)&=&\psi(n+1)-\psi(1),\\
 S_2 (n)&=&\zeta(2)-\psi'(n+1)=\zeta(2)-\zeta(2,n+1), \\
S_\alpha (n)&=&\zeta(\alpha)-\zeta(\alpha,n+1).
\ea
 \ba
\tilde{S}^{\pm}(n)&=&S_{-2,1}= -\frac{5}{8}\zeta(3) \mp \sum_{k=1}^{\infty}\frac{(-1)^k}{(k+n)^2}
\left(\psi(k+n+1)- \psi(1)\right).
   \ea
   \end{subequations}
For the $ S^{\pm}$ and $\tilde{S}$ series we  use the  notation given in
\cite{BK1999, KV2005}. Performing the analytic continuation from even $n$, $S^{+}_\alpha$, and from odd
$n$, $S^{-}_\alpha$ ( see for details \cite{KV2005}) one obtains
\begin{subequations}
 \label{eq:San}
 \ba \label{eq:S+}
&&\!\!\!\!\!\!\!\!\!\!\!\!S^{+}_\alpha(n/2)\!\to \!2^{\alpha-1}\left[S_\alpha(n)+S^{+}_{-\alpha}(n)\right]\!\!=
\!\!2^{\alpha-1}\left[S_\alpha(n)+ \zeta(\alpha)-\Phi(-1,\alpha,n+1)\right]-  \zeta(\alpha), \\
&&\!\!\!\!\!\!\!\!\!\!\!\!S^{-}_\alpha(n/2)\!\to \!2^{\alpha-1}\left[S_\alpha(n)+S^{-}_{-\alpha}(n)\right]\!\!=
\!\!2^{\alpha-1}\left[S_\alpha(n)+ \zeta(\alpha)+\Phi(-1,\alpha,n+1)\right]-  \zeta(\alpha),\label{eq:S-}
  \ea
  \end{subequations}
where $\Phi(z,\alpha,v)$ is the Lerch transcendent function \cite{Erde53}.
 The expressions on the r.h.s. of Eqs.(\ref{eq:San}) are now \textit{analytic functions in
 both variables $\alpha, n$} -- this is a new result.

\section{QCD \lowercase{evolution of Moments up to} NLO}
\label{app1}
The PDFs are the nonsinglet $f_\text{NS}(x,Q^2)$ and singlet $f_\text{S}(x,Q^2)$
parton distribution functions, %  $G(x,Q^2)$ -- gluon distribution
\ba
f_{\rm NS}(x,Q^2)&=&u_v(x,Q^2)-d_v(x,Q^2),
\\
f_{\rm S}(x,Q^2)& = &u_v(x,Q^2)+d_v(x,Q^2) +S(x,Q^2) \equiv V(x,Q^2)+S(x,Q^2),
\label{nss}
\ea
whereas $V(x,Q^2)$ is the distribution of valence quarks
and $S(x,Q^2)$ is the sea quark distribution.
More generally, the NS PDF is a
combination of the forms $u - d$ and $\bar{d} - \bar{u}$ but for our consideration we focus on
the nucleon scattering provided by combination (B1).
The moments representation for PDFs is defined as
\ba
f_{\rm NS}(n,Q^2)&=& \int_0^{1}dx x^{n-1}f_{\rm NS}(x,Q^2), \label{eq:NSmpdf}\\
f_{\rm S}(n,Q^2)&=&\int_0^{1}dx x^{n-1}f_{\rm S}(x,Q^2). \label{eq:mpdf}%\\
%G(n,Q^2)&=&\int_0^{1}dx x^{n}G(x,Q^2).
\ea
The moments $M_\text{NS}(n,\mu^2)$ for the structure function $F_\text{NS}(x,\mu^2)$ follow from
the Mellin convolution $F\stackrel{def}{=}C \ast f$,
\begin{subequations}
 \label{eq:Func-Mom}
\ba
F_\text{NS}(z,\mu^2)&=&\left(C_\text{NS} \ast f_\text{NS}\right)(z,\mu^2)\equiv\int_{0}^{1}C_\text{NS}(y,a_s)
f_\text{NS}(x,\mu^2)~\delta(z-x\cdot y) dy\,dx\, , \\
M_\text{NS}(n,\mu^2)&=& C_\text{NS}(n,a_s(\mu^2)) \cdot f_\text{NS}(n,\mu^2) \,.
\ea
 \end{subequations}
Here $C_\text{NS}(x,a_s)$ is the nonsinglet coefficient function of the process
that can be presented as the perturbation series
$C_\text{NS}(x,a_s) = 1 + a_s(Q^2) C_\text{NS}^{(1)}(x) + O(a_s^2)$;
$C_\text{NS}^{(1)}(n)$ in Appendix \ref{app0} is the moment of the $C_\text{NS}^{(1)}(x)$.
The QCD evolution of the moments $M_\text{NS}$ up to NLO of is given by (see Ref~\cite{YndBook})
\begin{subequations}
 \ba
M_\text{NS}(n,Q^2) =
%\nonumber\\
%& = &
\frac{1+C_\text{NS}^{(1)}(n) a_s(Q^2)}
{1+C_\text{NS}^{(1)}(n) a_s(Q_0^2)} \left(\frac{1+(\beta_1/\beta_0) a_s(Q^2)}
{1+(\beta_1/\beta_0) a_s(Q_0^2)} \right)^{p(n)}
\left[\frac{a_s(Q^2)}{a_s(Q_0^2)}\right]^{d_\text{NS}(n)}
\nonumber\\
 \times M_\text{NS}(n,Q_0^2),
 \label{mns}
% \nonumber
%& \simeq &
% \frac{1+\left(C_\text{NS}^{(1)}(n)+\frac{\beta_1}{\beta_0}p(n)
%\right)\alpha_s(Q^2)/4\pi}{1+\left(C_\text{NS}^{(1)}(n)+\frac{\beta_1}{\beta_0}p(n) \right)\alpha_s(Q_0^2)/4\pi}
%\left[\frac{\alpha_s(Q^2)}{\alpha_s(Q_0^2)}\right]^{\delta_\text{NS}(n)}
% \mu_\text{NS}(n,Q_0^2)
\ea
 where
\begin{equation}
M_\text{NS}(n,Q_0^2)=\left(1+C_\text{NS}^{(1)}(n) a_s(Q_0^2) \right)f_\text{NS}(n,Q_0^2)\, ,
\end{equation}
and:
\be
d_\text{NS}(n)=\gamma_\text{NS}^{(0)}(n)/2\beta_0,
~~p(n) = \frac{1}{2} \left(\frac{\gamma_\text{NS}^{(1)}(n)}{\beta_1}-\frac{\gamma_\text{NS}^{(0)}(n)}{\beta_0}
\right)\,.
\ee
\end{subequations}
The coefficients of anomalous dimension in LO and NLO and the coefficient function
in NLO are given in Eqs.(\ref{eq:AD-LO-NLO}), Appendix \ref{app0}.
In the case of the nucleon structure function $F_2(x,Q^2)$, one needs to take into account only even
values of $n$ in the NLO anomalous dimension.

\section{A\lowercase{ccuracy of the} J\lowercase{acobi} P\lowercase{olynomial Method}}
\label{app3}
The accuracy of the evaluation of the structure functions depends on the method we use;
therefore, it is indispensable to verify it in our approach.
The Jacobi Polynomial method promises us a good enough accuracy for the evolution,
as was shown in previous works (see~\cite{Krivokhizhin}).

\begin{figure}[ht] %\unitlength=1mm
\centering\includegraphics[width=150mm]{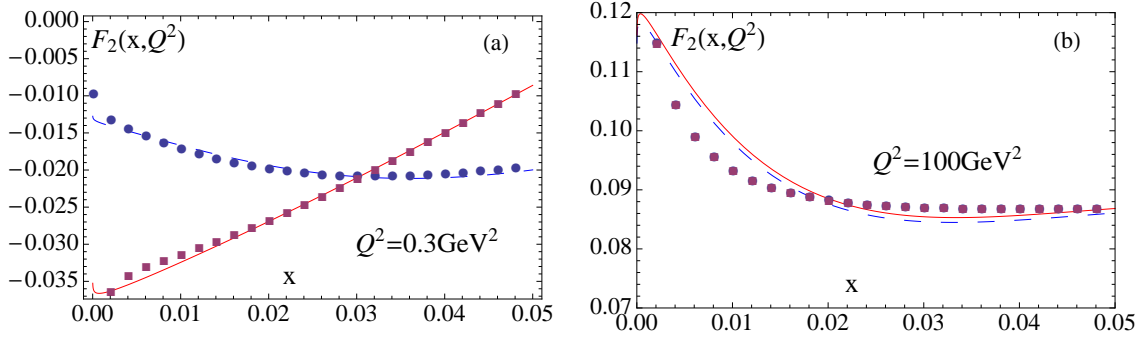}
\vspace{-0.4cm}
\caption{\footnotesize Nonsinglet SF $F_2(x)$ vs  $x$ at LO , at energy scale (a)
$Q^2=0.3$~GeV$^2$, and (b) $Q^2=100$~GeV$^2$.
The solid (red) line represents the FAPT outcome and the dashed (blue) line the  pQCD one in the JP method.
The thick squares (red) and spheres (blue) represent the result of the ``exact'' numerical Inverse Mellin transform.}
\label{figsf4}
 \end{figure}

This method is applied directly to the terms of the Bjorken variable $x$,
but it affects the $Q^2$-dependence indirectly.
Therefore, it is necessary to confirm the $x$-range applicability of the JP method.
To this end, we compare the results of the JP
approach with  the ``exact'' numerical calculations of  inverse Mellin moments following
Eq.~(\ref{exInvM}) but only
in the one-loop approximation due to technical limitations.
\begin{figure}[hb] %\unitlength=1mm
\centering\includegraphics[width=150mm]{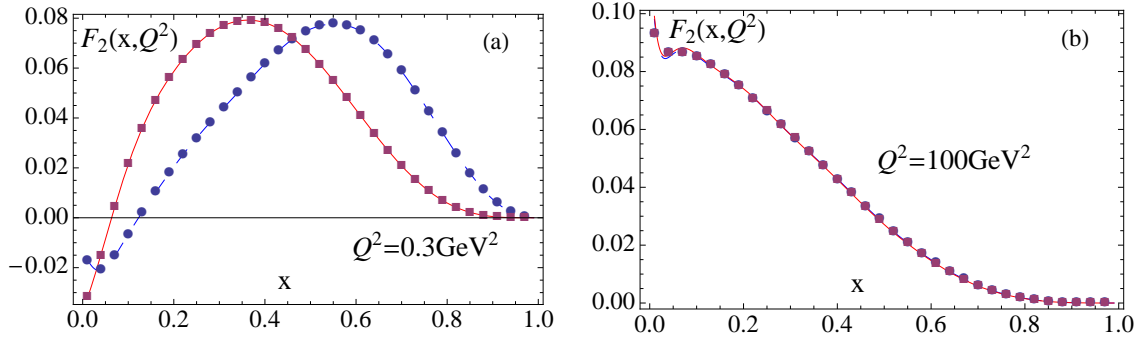}
\vspace{-0.4cm}
\caption{\footnotesize Nonsinglet SF $F_2(x)$ vs  $x$ in LO (a), at the energy scale
$Q^2=0.3$~GeV$^2$ and (b) $Q^2=100$~GeV$^2$.
The solid (red) line represents the FAPT result and the dashed (blue) line the pQCD one in the JP method.
The thick squares (red) and spheres (blue) represent the result of the ``exact'' numerical Inverse Mellin transform.}
\label{figsf3}
 \end{figure}
The comparison of these two results in Fig.~\ref{figsf3} demonstrates a very good accuracy.
So, in order to clarify it, we perform a zoom in $x$,
 going to a lower $x$-region ($\sim~10^{-2}$).
 We see in Fig.~\ref{figsf4} that the JP method gradually loses
 precision starting at $x < 0.02$. Also, we can see that in this range, the difference between these two
 methods reaches $5$\%.

\section{A\lowercase{ccuracy of the rational approximation}}
\label{app2}
Here we investigate the accuracy of the rational approximation for the two-loop evolution
factor

\be
m(n,Q^2)\equiv \left(\frac{1+(\beta_1/\beta_0) a_s(Q^2)}
{1+(\beta_1/\beta_0) a_s(Q_0^2)} \right)^{p(n)},
\label{approxM}
\ee
 in Eq.(\ref{appMpQCD}) for pQCD,  and Eq.(\ref{mnsapt}) for FAPT, respectively.
The expansion of the factor in power series up to NLO leads
\begin{subequations}
 \label{eq:C2}
\ba
m^{(1)}_\text{pQCD}(n,Q^2)&\simeq& \frac{1+(\beta_1/\beta_0)p(n) a_s(Q^2)}
{1+(\beta_1/\beta_0)p(n) a_s(Q_0^2)}, \\
m^{(2)}_\text{pQCD}(n,Q^2)&\simeq& \frac{1+(\beta_1/\beta_0)p(n)
a_s(Q^2)+(\beta_1^2/2\beta_0^2)p(n)(p(n)-1)a_s^2(Q^2)}
{1+(\beta_1/\beta_0)p(n) a_s(Q_0^2)+(\beta_1^2/2\beta_0^2)p(n)(p(n)-1)a_s^2(Q_0^2)},
\ea
 \end{subequations}
where $m^{(1)}(n,Q^2)$ and $m^{(2)}(n,Q^2)$ represent the approximation up to $\mathcal{O}(a_s)$ and
$\mathcal{O}(a_s^2)$, respectively. The corresponding ``FAPT form'' of (\ref{eq:C2}) is given by
\begin{subequations}
 \label{eq:C3}
\ba
m^{(1)}_\text{FAPT}(n,Q^2)&\simeq& \frac{1+(\beta_1/\beta_0)p(n) \A_1(Q^2)}
{1+(\beta_1/\beta_0)p(n) \A_1(Q_0^2)}, \\
m^{(2)}_\text{FAPT}(n,Q^2)&\simeq& \frac{1+(\beta_1/\beta_0)p(n)
\A_1(Q^2)+(\beta_1^2/2\beta_0^2)p(n)(p(n)-1)\A_2(Q^2)}
{1+(\beta_1/\beta_0)p(n)\A_1(Q_0^2)+(\beta_1^2/2\beta_0^2)p(n)(p(n)-1)\A_2(Q_0^2)}.
\ea
 \end{subequations}
Combing the approximations in Eqs. (\ref{eq:C2},~\ref{eq:C3}) in quantity
$\Ds \Delta m^{(12)}=|m^{(1)}-m^{(2)}|/m^{(1)}$
we obtain the accuracy better than $1\%$ for any $n \leq 13$
(since JP expansion contains only 13 terms for good approximation),
 in both cases of pQCD and FAPT for two different ranges of energy, i.e., low
$Q^2\sim1\,{\rm GeV}^2$ and high $Q^2\sim100\,{\rm GeV}^2$.
The results collected in Table \ref{Table1}
\begin{table}[ht]
\caption{\footnotesize The accuracy in per cent of the difference of the approximations
$\Ds \Delta m^{(12)}=\frac{|m^{(1)}-m^{(2)}|}{m^{(1)}}$
%of~(\ref{approxM}) up to $\mathcal{O}(a_s^1)$ and $\mathcal{O}(a_s^2)$
for pQCD: $\Delta m_{{\rm pQCD}}^{(12)}$ and for
FAPT: $\Delta m_{{\rm FAPT}}^{(12)}$. The results are presented in two ranges of $Q^2$:
low $Q^2\sim 1~{\rm GeV}^2$ and high $Q^2\sim 100~{\rm GeV}^2$.}
\label{Table1}
\begin{ruledtabular}
\begin{center}
\centering
\begin{tabular}{lllllllll}
%\hline
$n$ & 2 & 4 & 6 & 8 & 10 & 12
\\
\hline
$\Delta m_{{\rm pQCD}}^{(12)} \%$&\textbf{0.74}/ &$\bm{0.49}$/&$\bm{0.26}$ / &$\bm{0.06}$/&$\bm{0.12}$/ &$\bm{0.3}$
\\
%\hline
$Q^2\sim \bm{1}~{\rm GeV^2}$/$100~{\rm GeV^2}$ & 0.65 & 0.45 & 0.25 & 0.06 & 0.12 & 0.3 \\
\hline
$\Delta m_{{\rm FAPT}}^{(12)} \%$ & \textbf{0.03} / & $\bm{0.02}$ / & $\bm{0.01}$ / & $\bm{0.00}$ / & $\bm{0.01}$ /
& $\bm{0.01}$ \\
%\hline
$Q^2\sim \bm{1}~{\rm GeV^2}$/$100~{\rm GeV^2}$ & 0.18 & 0.13 & 0.07 & 0.02 & 0.04 & 0.09 \\
%\hline
\end{tabular}
\end{center}
\end{ruledtabular}
\end{table}
demonstrate that for both ranges of energy FAPT has a better convergence than pQCD;
even more,  the accuracy is improved for FAPT at low $Q^2\sim1~{\rm GeV}^2$
(really, pQCD must be worse but we have a low starting point $Q_0^2=1~{\rm GeV}^2$).
The strong hierarchy of FAPT couplings, $|\A_{\nu+1}^{\rm (FAPT)}(Q^{2})|
\ll |\A_{\nu}^{\rm (FAPT)}(Q^{2})|$, remains valid even at very low $|Q^2|$,
cf.~\cite{Bakulev}.


\begin{thebibliography}{99}
\bibitem{ShS}
%\cite{Shirkov:1996cd}
  D.~V.~Shirkov and I.~L.~Solovtsov,
  %``Analytic QCD running coupling with finite IR behaviour
%and universal ${\bar \alpha}_s(0)$ value,''
  [hep-ph/9604363];
  %%CITATION = HEP-PH 9604363;%%
%\cite{Shirkov:1997wi}
%  D.~V.~Shirkov and I.~L.~Solovtsov,
  %``Analytic model for the QCD running coupling with universal
%alpha(s)-bar(0) value,''
  Phys.~Rev.~Lett.~{\bf 79}, 1209 (1997)
  [hep-ph/9704333].
  %%CITATION = PRLTA,79,1209;%%
%1

\bibitem{MS}
  K.~A.~Milton, I.~L.~Solovtsov
  %``Analytic perturbation theory and inclusive tau decay,''
 Phys.\ Rev.\  D {\bf 55}, 5295 (1997)
  [hep-ph/9611438].
 %%CITATION = HEP-PH/9611438;%%
%2

\bibitem{MSS}
  K.~A.~Milton, I.~L.~Solovtsov and O.~P.~Solovtsova,
  %``Analytic perturbation theory and inclusive tau decay,''
  Phys.\ Lett.\ B {\bf 415}, 104 (1997)
  [arXiv:hep-ph/9706409].
   %%CITATION = PHLTA,B415,104;%%
%3
\bibitem{MSa}
  K.~A.~Milton, O.~P.~Solovtsova
 %"Analytic perturbation theory: A New approach to the analytic continuation of the strong coupling constant
 %alpha-s into the timelike region
 Phys.\ Rev.\  D {\bf 57}, 5402 (1998)
  [hep-ph/9710316].
 %%CITATION = HEP-PH/9710316;%%
%4

\bibitem{Sh1}
%\cite{Shirkov:2000qv}
  D.~V.~Shirkov,
  %``Analytic perturbation theory for QCD observables,''
  Theor.\ Math.\ Phys.\  {\bf 127}, 409 (2001)
  [hep-ph/0012283]; \textbf{119},  438  (1999).
  %%CITATION = TMPHA,127,409;%%
  %%CITATION = HEP-TH 9810246;%%
%5


%\cite{Karanikas:2001cs}
\bibitem{KS06}
  A.~I.~Karanikas and N.~G.~Stefanis,
  %``Analyticity and power corrections in hard scattering hadronic functions,''
  Phys.\ Lett.\ B {\bf 504}, 225 (2001)
  [Erratum-ibid.\ B {\bf 636}, 330 (2006)]
  [hep-ph/0101031].
  %%CITATION = HEP-PH/0101031;%%
  %58 citations counted in INSPIRE as of 18 mar 2015


\bibitem{BMS1}
%\bibitem{Bakulev:2005gw}
  A.~P.~Bakulev, S.~V.~Mikhailov and N.~G.~Stefanis,
  %``QCD analytic perturbation theory: From integer powers to any power of the
  %running coupling,''
  Phys.\ Rev.\  D {\bf 72}, 074014 (2005)
  [Erratum-ibid.\  D {\bf 72}, 119908 (2005)]
  [hep-ph/0506311].
  %%CITATION = PHRVA,D72,074014;%%
%6

\bibitem{BMS2}
%\bibitem{Bakulev:2006ex}
  A.~P.~Bakulev, S.~V.~Mikhailov and N.~G.~Stefanis,
  %``Fractional Analytic Perturbation Theory in Minkowski space and application
  %to Higgs boson decay into a b anti-b pair,''
  Phys.\ Rev.\  D {\bf 75}, 056005 (2007)
  [Erratum-ibid.\  D {\bf 77}, 079901 (2008)]
  [hep-ph/0607040].
  %%CITATION = PHRVA,D75,056005;%%
%7

\bibitem{BMS3}
%\bibitem{Bakulev:2010gm}
  A.~P.~Bakulev, S.~V.~Mikhailov, N.~G.~Stefanis,
  %``Higher-order QCD perturbation theory in different schemes: From FOPT to
  %CIPT to FAPT,''
  JHEP {\bf 1006}, 085 (2010)
  [arXiv:1004.4125 ].
  %%CITATION = JHEPA,1006,085;%%
%8

\bibitem{Bakulev}
A.~P.~Bakulev,
  %``Global Fractional Analytic Perturbation Theory in QCD with Selected
  %Applications,''
  Phys.\ Part.\ Nucl.\  {\bf 40}, 715 (2009)
  [arXiv:0805.0829 ] (arXiv preprint in Russian);
  %%CITATION = ARXIV:0805.0829;%%
%\bibitem{Stefanis:2009kv}
  N.~G.~Stefanis,
  %``Taming Landau singularities in QCD perturbation theory: The Analytic approach,''
  Phys.\ Part.\ Nucl.\  {\bf 44}, 494 (2013)
  [arXiv:0902.4805].
  %%CITATION = ARXIV:0902.4805;%%
%\bibitem{Bakulev:2011hp}
%  A.~P.~Bakulev and D.~V.~Shirkov,
%``Inevitability and Importance of Non-Perturbative Elements in Quantum Field Theory,''
%  arXiv:1102.2380;
%  %%CITATION = ARXIV:1102.2380;%%
%9

%\cite{Sidorov:2013aza}
\bibitem{Sidorov2013}
  A.~V.~Sidorov and O.~P.~Solovtsova,
  %``The QCD analysis of the combined set for the F_3 structure function data based on the analytic approach,''
  Mod. Phys. Lett. \textbf{A~29} (2014) 1450194, [arXiv:1407.6858];
  %%CITATION = ARXIV:1407.6858;%%
  Nonlin.\ Phenom.\ Complex Syst.\  {\bf 16}, 397 (2013),
  [arXiv:1312.3082 ].
  %%CITATION = ARXIV:1312.3082;%%
%10

 %\cite{Shirkov:2006gv}
\bibitem{Shirkov:2006gv}
  D.~V.~Shirkov and I.~L.~Solovtsov,
  %``Ten years of the analytic perturbation theory in QCD,''
  Theor.\ Math.\ Phys.\  {\bf 150}, 132 (2007)
  [hep-ph/0611229].
  %%CITATION = TMPHA,150,132;%%
%11

\bibitem{Nesterenko}
%\cite{Nesterenko:1999dx}
  A.~V.~Nesterenko,
  %``Quark antiquark potential in the analytic approach to QCD,''
  Phys.\ Rev.\ D {\bf 62}, 094028 (2000);
  %%CITATION = HEP-PH 9912351;%%
%\cite{Nesterenko:2001st}
%  A.~V.~Nesterenko,
  %``New analytic running coupling in spacelike and timelike regions,''
  Phys.\ Rev.\ D {\bf 64}, 116009 (2001);
  %%CITATION = HEP-PH 0102124;%%
%\cite{Nesterenko:2003xb}
%  A.~V.~Nesterenko,
  %``Analytic invariant charge in QCD,''
  Int.\ J.\ Mod.\ Phys.\ A {\bf 18}, 5475 (2003).
  %%CITATION = HEP-PH 0308288;%%
%\cite{Nesterenko:2004tg}
%12

\bibitem{Nesterenko2}
  A.~V.~Nesterenko and J.~Papavassiliou,
  %``The massive analytic invariant charge in QCD,''
  Phys.\ Rev.\ D {\bf 71}, 016009 (2005);
  %%CITATION = HEP-PH 0410406;%%
%\cite{Aguilar:2005sb}
  A.~C.~Aguilar, A.~V.~Nesterenko and J.~Papavassiliou,
  %``Infrared enhanced analytic coupling and chiral symmetry breaking in  QCD,''
  J.\ Phys.\ G {\bf 31}, 997 (2005).
  %%CITATION = HEP-PH 0504195;%%
%\bibitem{Nesterenko:2005wh}
%  A.~V.~Nesterenko and J.~Papavassiliou,
  %``Infrared behavior of the Adler function from a novel dispersion
  %relation,''
  J.\ Phys.\ G {\bf 32}, 1025 (2006)
  [hep-ph/0511215];
  %%CITATION = JPHGB,G32,1025;%%
%\cite{Nesterenko:2007fm}
%\bibitem{Nesterenko:2007fm}
  A.~V.~Nesterenko,
  %``Adler function in the analytic approach to QCD,''
  arXiv:0710.5878.
  %%CITATION = ARXIV:0710.5878;%%
%13

%\cite{Alekseev:2005he}
\bibitem{Alekseev:2005he}
  A.~I.~Alekseev,
  %``Synthetic running coupling of QCD,''
  Few Body Syst.\  {\bf 40}, 57 (2006)
  [hep-ph/0503242].
  %%CITATION = FBSYE,40,57;%%
%14

%\cite{Srivastava:2001ts}
\bibitem{Srivastava:2001ts}
  Y.~Srivastava, S.~Pacetti, G.~Pancheri and A.~Widom,
  %``Dispersive techniques for alpha(s), R(had) and instability of the
  %perturbative vacuum,''
%in {\it Proc. of the $e^+ e^-$ Physics at Intermediate Energies Conference } ed. Diego Bettoni,
{\it In the Proceedings of $e^+ e^-$ Physics at Intermediate Energies, SLAC, Stanford, CA, USA, 30 April - 2 May
2001, pp T19}
  [hep-ph/0106005].
  %%CITATION = ECONF,C010430,T19;%%
%15

%\cite{Webber:1998um}
\bibitem{Webber:1998um}
  B.~R.~Webber,
  %``{QCD} power corrections from a simple model for the running coupling,''
  JHEP {\bf 9810}, 012 (1998)
  [hep-ph/9805484].
  %%CITATION = JHEPA,9810,012;%%
%16

\bibitem{CV1}
%\bibitem{Cvetic:2006mk}
  G.~Cveti\v c and C.~Valenzuela,
  %``An approach for evaluation of observables in analytic versions of QCD,''
  J.\ Phys.\ G {\bf 32}, L27 (2006)
  [hep-ph/0601050].
  %%CITATION = JPHGB,G32,L27;%%
%17

\bibitem{CV2}
%\bibitem{Cvetic:2006gc}
  G.~Cveti\v c and C.~Valenzuela,
  %``Various versions of analytic QCD and skeleton-motivated evaluation of
  %observables,''
  Phys.\ Rev.\  D {\bf 74}, 114030 (2006)
  [hep-ph/0608256].
  %%CITATION = PHRVA,D74,114030;%%
%18

%\cite{Prosperi:2006hx}
\bibitem{Prosperi:2006hx}
  G.~M.~Prosperi, M.~Raciti and C.~Simolo,
  %``On the running coupling constant in QCD,''
  Prog.\ Part.\ Nucl.\ Phys.\  {\bf 58}, 387 (2007)
  [hep-ph/0607209].
  %%CITATION = PPNPD,58,387;%%
%19

%\cite{Cvetic:2008bn}
\bibitem{Cvetic:2008bn}
  G.~Cveti\v c and C.~Valenzuela,
  %``Analytic QCD - a short review,''
  Braz.\ J.\ Phys.\  {\bf 38}, 371 (2008)
  [arXiv:0804.0872 ].
  %%CITATION = BJPHE,38,371;%%
%20

%\cite{Belyakova:2010iw}
\bibitem{Belyakova:2010iw}
  Y.~O.~Belyakova and A.~V.~Nesterenko,
  %``A nonperturbative model for the strong running coupling within potential approach,''
  Int. J. Mod. Phys.\ A {\bf 26}, 981 (2011)
  [arXiv:1011.1148].
  %%CITATION = ARXIV:1011.1148;%%
%21

\bibitem{CKV1}
  G.~Cveti\v{c}, R.~K\"ogerler and C.~Valenzuela,
  %``Analytic QCD coupling with no power terms in UV regime,''
  J.\ Phys.\ G {\bf 37}, 075001 (2010)
  [arXiv:0912.2466].
  %%CITATION = JPHGB,G37,075001;%%
%22

\bibitem{CKV2}
%\bibitem{Cvetic:2010di}
G.~Cveti\v{c}, R.~K\"ogerler and C.~Valenzuela,
  %``Reconciling the analytic QCD with the ITEP operator product expansion philosophy,''
  Phys. Rev. {\bf D}  82, 114004 (2010)
  [arXiv:1006.4199].
  %%CITATION = ARXIV:1006.4199;%%
%23

%\cite{Contreras:2014aka}
\bibitem{CCKKO}
  C.~Contreras, G.~Cveti\v{c}, R.~ K\"ogerler, P.~Kroger and O.~Orellana,
  %``Perturbative QCD in acceptable schemes with holomorphic coupling,''
  arXiv:1405.5815 .
  %%CITATION = ARXIV:1405.5815;%%

\bibitem{CCEMAyala}
  C.~Ayala, C.~Contreras and G.~Cveti\v{c},
  %``Extended analytic QCD model with perturbative QCD behavior at high momenta,''
  Phys.\ Rev.\ D {\bf 85}, 114043 (2012)
  [arXiv:1203.6897 ].
  %%CITATION = ARXIV:1203.6897;%%


%\cite{naiveF2}
\bibitem{naiveF2}
  A.~V.~Kotikov, V.~G.~Krivokhizhin and B.~G.~Shaikhatdenov,
  %``Analytic and 'frozen' QCD coupling constants up to NNLO from DIS data,''
  Phys.\ Atom.\ Nucl.\  {\bf 75}, 507 (2012)
  [arXiv:1008.0545 [hep-ph]];
  %%CITATION = ARXIV:1008.0545;%%
  %24 citations counted in INSPIRE as of 24 Apr 2015
%\cite{Cvetic:2009kw}
%\bibitem{Cvetic:2009kw}
  G.~Cveti\v{c}, A.~Y.~Illarionov, B.~A.~Kniehl and A.~V.~Kotikov,
  %``Small-x behavior of the structure function F(2) and its slope partial ln F(2) / partial ln(1/x) for 'frozen' and analytic strong-coupling constants,''
  Phys.\ Lett.\ B {\bf 679}, 350 (2009)
  [arXiv:0906.1925 [hep-ph]].
  %%CITATION = ARXIV:0906.1925;%%
  %30 citations counted in INSPIRE as of 24 Apr 2015

%\cite{BjSR}
\bibitem{BjSR}
  R.~S.~Pasechnik, D.~V.~Shirkov and O.~V.~Teryaev,
  %``Bjorken Sum Rule and pQCD frontier on the move,''
  Phys.\ Rev.\ D {\bf 78}, 071902 (2008)
  [arXiv:0808.0066 [hep-ph]];
  %%CITATION = ARXIV:0808.0066;%%
  %41 citations counted in INSPIRE as of 24 Apr 2015
%\cite{Pasechnik:2009yc}
%\bibitem{Pasechnik:2009yc}
  R.~S.~Pasechnik, D.~V.~Shirkov, O.~V.~Teryaev, O.~P.~Solovtsova and V.~L.~Khandramai,
  %``Nucleon spin structure and pQCD frontier on the move,''
  Phys.\ Rev.\ D {\bf 81}, 016010 (2010)
  [arXiv:0911.3297 [hep-ph]];
  %%CITATION = ARXIV:0911.3297;%%
  %30 citations counted in INSPIRE as of 24 Apr 2015
%\cite{Khandramai:2011zd}
%\bibitem{Khandramai:2011zd}
  V.~L.~Khandramai, R.~S.~Pasechnik, D.~V.~Shirkov, O.~P.~Solovtsova and O.~V.~Teryaev,
  %``Four-loop QCD analysis of the Bjorken sum rule vs data,''
  Phys.\ Lett.\ B {\bf 706}, 340 (2012)
  [arXiv:1106.6352].
  %%CITATION = ARXIV:1106.6352;%%
  %19 citations counted in INSPIRE as of 24 Apr 2015

\bibitem{OT2013}
O.~Teryaev, Nucl.Phys.Proc.Suppl. 245 (2013) 195 [arXiv:1309.1985 ].
%%CITATION = ARXIV:1309.1985 [hep-ph];%%

%\cite{Allendes:2014fua}
\bibitem{GluonProp}
  P.~Allendes, C.~Ayala and G.~Cveti\v{c},
  %``Gluon Propagator in Fractional Analytic Perturbation Theory,''
  Phys.\ Rev.\ D {\bf 89}, 054016 (2014)
  [arXiv:1401.1192].
  %%CITATION = ARXIV:1401.1192;%%
  %5 citations counted in INSPIRE as of 24 Apr 2015


%\cite{Gribov:1972ri}
\bibitem{dglap}
  V.~N.~Gribov and L.~N.~Lipatov,
  %``Deep inelastic e p scattering in perturbation theory,''
  Sov.\ J.\ Nucl.\ Phys.\  {\bf 15}, 438 (1972)
  [Yad.\ Fiz.\  {\bf 15}, 781 (1972)];
  %%CITATION = SJNCA,15,438;%%
%\cite{Lipatov:1974qm}
  L.~N.~Lipatov,
  %``The parton model and perturbation theory,''
  Sov.\ J.\ Nucl.\ Phys.\  {\bf 20}, 94 (1975)
  [Yad.\ Fiz.\  {\bf 20}, 181 (1974)];
  %%CITATION = SJNCA,20,94;%%
%\cite{Altarelli:1977zs}
  G.~Altarelli and G.~Parisi,
  %``Asymptotic Freedom in Parton Language,''
  Nucl.\ Phys.\ B {\bf 126}, 298 (1977);
  %%CITATION = NUPHA,B126,298;%%
%\cite{Dokshitzer:1977sg}
  Y.~L.~Dokshitzer,
  %``Calculation of the Structure Functions for Deep Inelastic Scattering and e+ e- Annihilation by Perturbation
  %Theory in Quantum Chromodynamics.,''
  Sov.\ Phys.\ JETP {\bf 46}, 641 (1977)
  [Zh.\ Eksp.\ Teor.\ Fiz.\  {\bf 73}, 1216 (1977)].
  %%CITATION = SPHJA,46,641;%%

%\cite{Krivokhizhin:1987rz}
\bibitem{Krivokhizhin}
V. G. Krivokhizhin et al.,
  %``Qcd Analysis Of Singlet Structure Functions Using Jacobi Polynomials: The Description Of The Method,''
  Z.\ Phys.\ C {\bf 36}, 51 (1987);
  %%CITATION = ZEPYA,C36,51;%%
  %\cite{Krivokhizhin:1990ct}
  %V.~G.~Krivokhizhin, S.~P.~Kurlovich, R.~Lednicky, S.~Nemecek, V.~V.~Sanadze, I.~A.~Savin, A.~V.~Sidorov and
 V. G. Krivokhizhin et al.,
  %``Next-to-leading order QCD analysis of structure functions with the help of Jacobi polynomials,''
  Z.\ Phys.\ C {\bf 48}, 347 (1990).
  %%CITATION = ZEPYA,C48,347;%%

  %\cite{Martin:2009iq}
\bibitem{Martin:2009iq}
  A.~D.~Martin, W.~J.~Stirling, R.~S.~Thorne and G.~Watt,
  %``Parton distributions for the LHC,''
  Eur.\ Phys.\ J.\ \textbf{C 63}, 189 (2009)
  [arXiv:0901.0002 ].
  %%CITATION = ARXIV:0901.0002;%%

  %\cite{Buras:1979yt}
\bibitem{Buras:1979yt}
  A.~J.~Buras,
  %``Asymptotic Freedom in Deep Inelastic Processes in the Leading Order and Beyond,''
  Rev.\ Mod.\ Phys.\  {\bf 52}, 199 (1980).
  %%CITATION = RMPHA,52,199;%%

%\cite{Krivokhizhin:2005pt}
\bibitem{Kotikov}
  V.~G.~Krivokhizhin and A.~V.~Kotikov,
  %``A systematic study of QCD coupling constant from deep-inelastic measurements,''
  Phys.\ Atom.\ Nucl.\  {\bf 68} (2005) 1873
   [Yad.\ Fiz.\  {\bf 68} (2005) 1935];
  %%CITATION = PANUE,68,1873;%%
  %\cite{Krivokhizhin:2009zz}
  V.~G.~Krivokhizhin and A.~V.~Kotikov,
  %``Functions of the nucleon structure and determination of the strong coupling constant,''
  Phys.\ Part.\ Nucl.\  {\bf 40}, 1059 (2009).
  %%CITATION = PPNUE,40,1059;%%


  %\cite{Bakulev:2012sm}
\bibitem{Bakulev:2012sm}
  A.~P.~Bakulev and V.~L.~Khandramai,
  %``FAPT: a Mathematica package for calculations in QCD Fractional Analytic Perturbation Theory,''
  Comput.Phys.Commun. \textbf{184}(2013) 1, 183
  arXiv:1204.2679.
  %%CITATION = ARXIV:1204.2679;%%


%\cite{Ayala:2014pha}
\bibitem{Ayala2014}
  C.~Ayala and G.~Cveti\v{c},
  %``anQCD: a Mathematica package for calculations in general analytic QCD models,''
  Comput.\ Phys.\ Commun.\  {\bf 190}, 182 (2015)
  [arXiv:1408.6868 [hep-ph]];
  %%CITATION = ARXIV:1408.6868;%%
  %4 citations counted in INSPIRE as of 18 Mar 2015
%\cite{Ayala:2014qea}
%\bibitem{Ayala:2014qea}
 % C.~Ayala and G.~Cvetic,
  %``Mathematica and Fortran programs for various analytic QCD couplings,''
  arXiv:1411.1581 [hep-ph].
  %%CITATION = ARXIV:1411.1581;%%
  %1 citations counted in INSPIRE as of 18 Mar 2015

  \bibitem{YndBook}
  F. J. Yndurain, The Theory of Quarks and Gluons Interactions (Fourth Edition)
 (Springer-Verlag, Berlin, 2006).

\bibitem{Erde53}
A. Erd\'elyi, W. Magnus, F. Oberhettinger and F. G. Tricomi (1953),
Higher Transcendental Functions. Vol. I, McGraw-Hill Book Company, Inc., New York-Toronto-London.

\bibitem{BK1999} J. Blumlein and S. Kurth,
%Harmonic sums and Mellin transforms up to two-loop order
Phys.\ Rev.\  {\bf D 60}, 014018 (1999);
\bibitem{KV2005} A.V. Kotikov and V. N. Velizhanin,
%Analytic continuation of the Mellin moments
%of deep inelastic structure functions
hep-ph/0501274


 %\cite{Floratos:1977au}
 %\bibitem{AD}
  E.~G.~Floratos, D.~A.~Ross and C.~T.~Sachrajda,
  %``Higher Order Effects in Asymptotically Free Gauge Theories: The Anomalous Dimensions of Wilson Operators,''
  Nucl.\ Phys.\ B {\bf 129}, 66 (1977)
  [Erratum-ibid.\ B {\bf 139}, 545 (1978)].
  %%CITATION = NUPHA,B129,66;%%
  %\cite{GonzalezArroyo:1979ng}
  A.~Gonzalez-Arroyo, C.~Lopez and F.~J.~Yndurain,
  %``Second Order Contributions To The Structure Functions In Deep Inelastic Scattering. Ii.    Comparison With
  %Experiment For The Nonsinglet Contributions To E, Mu Nucleon Scattering,''
  Nucl.\ Phys.\ B {\bf 159}, 512 (1979).
  %%CITATION = NUPHA,B159,512;%%
  %\cite{Curci:1980uw}
  G.~Curci, W.~Furmanski and R.~Petronzio,
  %``Evolution of Parton Densities Beyond Leading Order: The Nonsinglet Case,''
  Nucl.\ Phys.\ B {\bf 175}, 27 (1980).
  %%CITATION = NUPHA,B175,27;%%

\end{thebibliography}
\end{document}